\documentclass[pre]{revtex4-1}

\begin{document}
\title{Can we model DNA at the mesoscale ?\\
Comment on: {\it Fluctuations in the DNA double helix: A critical
  review.} }

\author{Michel Peyrard$^{1}$ and 
Thierry Dauxois$^{1}$
} 

\affiliation{
$^{1}$ Ecole 
Normale Sup\'erieure de Lyon, Laboratoire de Physique CNRS,
46 all\'ee d'Italie, 69364 Lyon Cedex 7, France
}

\date{\today}

\maketitle

\bigskip
This review includes a discussion on the theoretical approaches
intending to provide a quantitative analysis of DNA fluctuations, 
but, as this topic is
very broad, it concentrates on two particular cases: the two-state
Poland-Scheraga model \cite{PSmodel} and the Peyrard-Bishop-Dauxois (PBD)
mesoscopic approach which describes the fluctuations of a base pair in terms
of a single variable \cite{PBD}, 
the stretching of the pair. The main point of the
authors, emphasized in the conclusion, is that any mesoscopic model does
not make sense, and as molecular dynamics cannot reach the time scales
needed to investigate an event so rare as base-pair opening, only the
two-state model remains. We think that such a sharp judgment can be
questioned.

\smallskip
Discarding all molecular dynamics studies is certainly unjustified. It
is true that the simulations face difficulties owing to the very
large number of atoms that have to be studied, even for a short DNA
fragment, due to the solvent, and owing to the gap in time scales between
small vibrational motions and the full opening of a base pair. However,
molecular dynamics is more than brute force calculations of time
trajectories. It can be used for a clever sampling of the phase space
allowing a calculation of the free energy pathways for base pair openings,
taking into account the solvent and counterions \cite{Giudice2003}, giving
in turn precious information for modeling DNA at an intermediate scale.

\bigskip
In their assessment of mesoscopic models, the authors of the review
adopt the same kind of narrow view, which leads them to two fundamental
misunderstandings: first, as in the case of molecular dynamics, the
review puts emphasis on the dynamics, which is not at all the best way
to derive useful results from a model, and second the authors do not
seem to realize that, in a mesoscopic model, the ``potentials'' are
actually containing the missing microscopic degrees of freedom in some
effective way so that a statement such as ``the model considers DNA as
if it where in vacuum'' is completely wrong. Let us comment on these two
points.

\medskip
{\em Dynamics versus statistical physics.} In contrast to the two-state
model, for a mesoscopic model such as PBD it is easy to write equations
of motions. This does not mean that they have to be used to derive the
most useful results that can be obtained with such a model. Very few
experiments are actually able to access time-dependent properties. 
Most of them probe equilibrium
properties, such as the average opening at a given temperature (i.e.\
the melting curve) or local equilibrium probabilities.
Even the studies discussed in the review, which 
determined the time scale of
the opening of the base pairs, were actually based on chemical equilibria.
The theory of these measurements is obtained much more accurately with
statistical physics than by extracting probabilities from dynamical
calculations. For rare events a dynamical approach can get particularly
unreliable \cite{Comment-Bishop}. For the PBD model,
statistical physics calculations
can be made very efficiently because the model does not include long--range
interactions, in contrast to the entropic contribution of the
two-state model. Provided the sequence is properly
included in the parameters, i.e.\ one goes beyond the uniform model
described in the review, accurate melting profiles for long sequences
can easily be deduced from a statistical physics calculations \cite{NTh}
whereas getting them from dynamical simulations
involving several tens of thousands of base pairs would be hopeless.

\medskip
{\em Potentials in a mesoscopic model.} A mesoscopic model attempts to
describe the properties of a system by considering some degrees of
freedom only. This does not mean that the other degrees of freedom are
totally ignored. Even molecular dynamics at the atomic scale is a
``mesoscopic'' model because it studies the positions of the atoms without
specifying the state of the electrons as an ab-initio simulation would do. 
Nobody would think that molecular
dynamics describes nuclei without electrons! Indeed the potentials
implicitly depends on the electronic degrees of freedom. The same is
true, at a larger scale, in a model of DNA which only describes the
stretching of the base pairs. The other degrees of freedom, including
the effect of the solvent or ions, enter in the ``potential''
parameters. Actually the word ``potentials'' for the functions of the
simplified Hamiltonian is not appropriate. One should have in mind free
energies, or what is called ``potential of mean force'' in some
molecular dynamics approaches \cite{GarciaPMF}. This is why for instance
the ionicity of the solution is taken into account when the PBD model is
used to quantitatively describe DNA melting curves \cite{NTh}. Therefore
a mesoscopic model certainly does not describe ``DNA in vacuum'',
although good effective potentials may be difficult to
obtain. Basic considerations can be used to evaluate some general
features, such as the dissociation energy of the base pairs in the
presence of the counterions, but a systematic elimination of the
microscopic degrees of freedom is a hard task. This explains why the
model parameters have evolved from the first comparison with experiments
for short DNA chains \cite{Campa}, to more recent applications to long
complex sequences \cite{NTh}. 

\bigskip 

As effective potentials are complex,
their choice may depend on the problem of interest. To fit a melting
curve, an appropriate description of the energy levels for the closed and
open states and of the entropic effects, introduced through a nonlinear
stacking term, is sufficient. If one intends to use the model for
dynamical properties such as the evaluation of the open--state lifetime, 
the effective potential determination is more demanding. As
pointed in the review, it is true that the potentials 
used in the original version of the PBD model lead to completely incorrect
results for the lifetime of the open state if the model is used in dynamical
simulations. They have been modified for that purpose  \cite{PBdynamics}. 
Nevertheless it is important to realize that small amplitude
motions of DNA {\em are not overdamped} otherwise one would not observe
the Raman peaks characteristic of the vibration of the molecule. This is
because water does not easily penetrate inside the base pair stack
although it may form hydrogen bonds in the groves. Only when the
amplitude of the fluctuations becomes so large that the bases get
fully embedded in water, does the damping increase strongly, as stated in the
review. However other phenomena are even more important to increase the
lifetime of the open state:
on the one hand out-of-the-stack bases have more
degrees of freedom than stacked bases, 
i.e.\ more entropy and, on the other hand, they can
form hydrogen bonds with the solvent. Both effects lower the free energy
of the open state, leading to an effective barrier for reclosing. 
This barrier does
not significantly change the statistical properties of the model
although it makes the denaturation transition a bit sharper
\cite{PBdynamics}, but it drastically changes its dynamics. When it is
included 
the lifetime of the open state increases by several orders of
magnitude so that it approaches the observed values
\cite{PBdynamics}. Further improvements would still be needed
for a quantitative description of DNA dynamics, in particular the introduction
of an extended Langevin simulation method taking into account the
variation of the damping with the amplitude of the fluctuations.

\bigskip
In addition to modeling, this review on DNA fluctuations also discusses
experiments.  Although our comment focuses on modeling, we would
nevertheless like to point out that experimental studies of DNA fluctuations can
still reveal unexpected results as shown by a recent
investigation using Guanine radical chemistry that demonstrated
that the large fluctuations of an AT-rich region may affect the
configurations of bases up to about 10 base pairs away \cite{CuestaNAR} 
even at physiological temperatures. This should not be a surprise
because the experiments, discussed in the review, which confirmed the
very low opening probability of base pairs \cite{Gueron87}, also show
that the activation enthalpy for the opening, which stays roughly
constant in the $0-25^{\circ}\;$C range, changes significantly above
$25^{\circ}\;$C. A straightforward extrapolation of the data of Fig.~12
of [\onlinecite{Gueron87}b] suggests an increase of the opening rate
by a factor of 10 between
$25^{\circ}\;$C and $40^{\circ}\;$C. As a result the role of DNA
fluctuations could even be more important than suggested in the review,
particularly when specific sequences are concerned.

\bigskip
The main message that we would like to convey in this comment is that
there is not one good model and one bad model for DNA. The reality is
more complex and depends on the questions of interest. As pointed out in
the review, the two-state model is successful and its numerous empirical
parameters are now well under control, so that it is widely used. 
A mesoscopic model, such as the PBD model, attempts to build the
description from a microscopic view by lumping many degrees of freedom
in a few variables at a larger scale. The interest is that, at least in
principle, the parameters of its effective potentials can be
quantitatively derived, although it is not a trivial task. Results from
an exploration of the phase space by molecular dynamics could be used
\cite{Giudice2003}. When statistical physics is used, both the two-state
and the PBD model give similar melting curves for complex DNA
sequences. However when one is interested in some peculiar properties
that depend on the magnitude of the fluctuations, a model going
beyond a two-state picture is required. This is for
instance the case in the analysis of the temperature dependence of the
broadening of some Bragg peaks in neutron diffraction
\cite{DNA-neutrons}. The width of the peak associated to the base
stacking reflects the correlation length of the double helix.
It increases sharply at the denaturation transition, when the
stack breaks into short segments, destroying spatial coherence. 
However prior to the full denaturation, the
width of the peak also depends on the magnitude of the transverse
fluctuations of the base pairs, which can be estimated from a statistical
physics study of the PBD model \cite{DNA-neutrons}.

\bigskip
The PBD model poses interesting questions in nonlinear dynamics, and, as
noticed in the review, it is rather easy to handle by mathematicians or
physicists. This is why it had many ``followers'' (to use the word of
the review) and some of them extrapolated to biology conclusions drawn
in conditions which are not relevant for DNA (no thermal bath,
inadequate parameters). This has led to statements such as the
prediction of solitons moving in DNA, which are easy to criticize. However,
exactly as one would not reject cars because bad drivers have killed
people with them, one should not discard the modeling of DNA at a
mesoscale. It provides further insights and it is likely that the
future lies in a combination of models at various scales, as done recently
in an hybrid approach which mixed molecular dynamics and mesoscale
modeling \cite{Meyer} to study DNA flexibility.

%%%%%%%%%%%%%%%%%%%

\end{document}